\documentclass[a4paper,11pt]{article}
\usepackage{pos}
\usepackage{placeins}
\usepackage{braket}

\newcommand{\p}{^{\prime}}

\newcommand{\vet}{\boldsymbol}

\newcommand{\smatrix}{\begin{pmatrix}}
\newcommand{\cmatrix}{\end{pmatrix}}

\newcommand{\da}{^\dagger}

\title{Fits of finite-volume smeared spectral densities}
\ShortTitle{Fits of finite-volume smeared spectral densities}

\author[a]{Luigi Del Debbio}
\author*[a]{Alessandro Lupo}
\author[b]{Marco Panero}
\author[c]{Nazario Tantalo}

\affiliation[a]{Higgs Centre for Theoretical Physics, School of Physics \& Astronomy, The University of Edinburgh, Peter Guthrie Tait Road, Edinburgh EH9 3FD, United Kingdom}
\affiliation[b]{Department of Physics, University of Turin \& INFN, Turin\\
Via Pietro Giuria 1, I-20125 Turin, Italy}
\affiliation[c]{University and INFN of Roma Tor Vergata\\
	Via della Ricerca Scientifica 1, I-00133, Rome, Italy}

\emailAdd{alessandro.lupo@ed.ac.uk}

\abstract{Motivated by recent progress in the numerical inversion of the Laplace transform, we investigate applications of finite-volume smeared spectral densities. These include the tuning of operator smearing, and the study of the finite-volume spectrum.}

\FullConference{%
The 39th International Symposium on Lattice Field Theory,\\
8th-13th August, 2022,\\
Rheinische Friedrich-Wilhelms-Universität Bonn, Bonn, Germany
}

\begin{document}
\maketitle

\section{Introduction}
\label{sec:intro}
The quantization in the path-integral formalism in a Euclidean spacetime enables the application of Monte Carlo methods in order to estimate observables in gauge theories. The connection between Euclidean correlation functions and Minkowski, physical amplitudes is hindered by off-shell contributions that arise at large Euclidean time~\cite{Maiani:1990ca}. Nonetheless, the spectrum of the finite-volume Hamiltonian is connected to infinite-volume Minkowski amplitudes~\cite{Luscher:1985dn, Luscher:1986pf, Kim:2005gf}, which are therefore accessible from lattice calculations. For this reason, the computation of the finite-volume energy spectra of gauge theories is a task of primary importance. Another connection between Euclidean correlation functions and Minkowski amplitudes is provided by spectral densities. These are independent of the metric, and they allow the extraction of information without relying on large time separations, avoiding the problems of Ref.~\cite{Maiani:1990ca}. The limitation lies in the fact that the computation of spectral densities from lattice correlators is ill-posed. Nonetheless, regularisations to this problem exist~\cite{10.1111/j.1365-246X.1968.tb00216.x, Hansen:2019idp}, and the topic has been receiving increasing attention~\cite{Hansen:2017mnd, Gambino:2020crt, Bulava:2021fre, Gambino:2022dvu, Bulava:2019kbi, Bruno:2020kyl, Bailas:2020qmv}. Applications to inclusive decays can be found in Refs.~\cite{Hansen:2017mnd, Gambino:2020crt, Bulava:2021fre}, and interesting ideas for exclusive processes are in Ref.~\cite{Bulava:2019kbi}.

In this proceeding we take another turn. Most of the aforementioned references are designed in the perspective of the infinite-volume limit. Here, we use smeared spectral densities to study the finite-volume spectra. There are several motivations for looking into this direction. The energies are encoded into lattice correlators as functions of Euclidean time, and spectral densities contain the same information as functions of the energy, providing a different outlook on lattice data. While the extraction of a ground state from a correlation function often relies on its large-time behaviour, the information is mixed non-trivially in the spectral densities, which takes contributions from the correlator at each time. We will show that the ground state can be extracted from smeared spectral densities by performing non-linear fits to the smearing kernel. We will also mention applications for operator smearing. In order to show these ideas at a reasonable computational cost, we analyse in this proceeding synthetic data and correlation functions of mesons.

\section{Smeared spectral densities}
Our setup for the computation of smeared spectral densities, introduced in Ref.~\cite{Hansen:2019idp}, has been extensively discussed in Ref.~\cite{DelDebbio:2022qgu}. In this proceeding, we smear the spectral density with a Gaussian kernel, $\Delta_\sigma(E) = \exp{(-E^2/2\sigma^2)}/\sqrt{2\pi}\sigma$
\begin{equation}
    \rho_{L,\sigma}(E) = \int dE\p \Delta_\sigma(E-E\p) \, \rho_L(E\p) \; . 
\end{equation}
We begin from the two point function, at zero momentum, built with the interpolating field $O(\vet{x},t)$, which can be written as
\begin{equation}\label{eq:c_expanded_cosh}
    C_{L}(t) = \sum_n \left( e^{-tE_n(L)} + e^{(-T+t)E_n(L)} \right) \, \frac{\braket{0|O(0)|n}_L \braket{n|O\da(0)|0}_L}{2E_n(L)} \; .
\end{equation}
The smeared spectral density associated with such correlator is then
\begin{equation}\label{eq:smeared_spectraldens_explicit}
    \rho_{L, \sigma}(E) = \sum_n \frac{\braket{0|O(0)|n}_L \braket{n|O\da(0)|0}_L}{2E_n(L)}\, \Delta_\sigma \left( E-E_n(L) \right) \; .
\end{equation}
It is clear from these two equations that the information encoded in correlators and spectral densities is the same. The idea of fitting Eq.~\eqref{eq:c_expanded_cosh} to a sum of exponentials in order to obtain energies and matrix elements is well established. In Sec.~\ref{sec:pos:fits}, we will touch on the possibility to fit instead Eq.~\eqref{eq:smeared_spectraldens_explicit} to a sum of Gaussians. In the following, we will drop references to the volume $L$, which is understood to be finite.
\begin{figure}[tb]
    \centering
    \includegraphics[width=\textwidth]{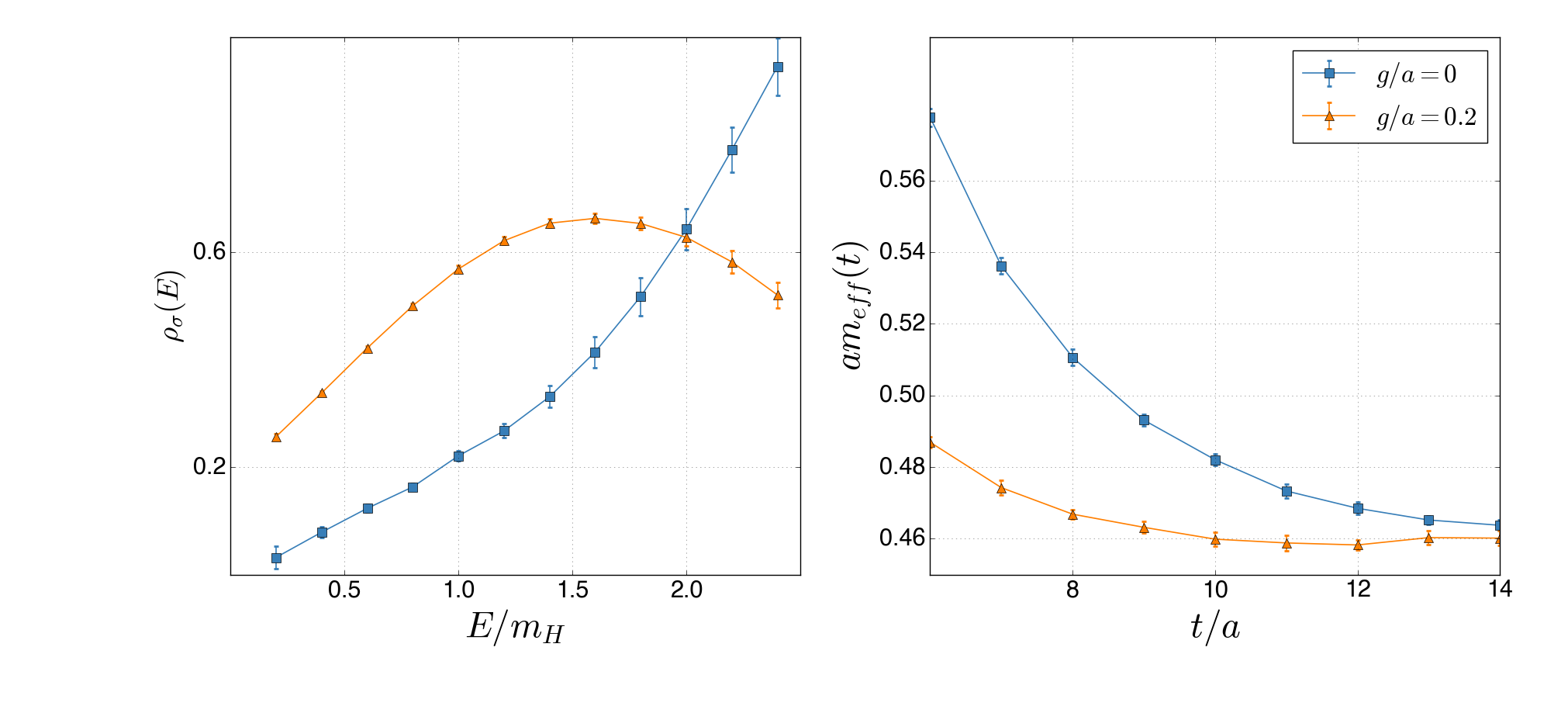}
    \caption{The effect of operator smearing is shown on the smeared spectral density (left panel) and on the correlator (right panel). Without operator smearing $(g=0)$, excited states dominate both signals, shown in blue. In the orange lines, the radius $g$ of Eq.~\eqref{eq:pos:opsm} has been tuned so that the interpolating operator mainly overlaps with those states $\ket{n}$ having energy $0\leq E_n \lesssim 2m_H$.}
    \label{fig:pos_smearing}
\end{figure}
\section{Operator smearing}
\label{sec:pos:smear}
Consider a local interpolating operator of the type $O(x) = \bar{\psi}(x) \Gamma \psi(x)$ that annihilates the hadron $H$. Such operator can have large overlap with excited states, hindering the study of the ground state $\ket{H}$. Operator smearing\footnote{Not to be confused with the smearing of the spectral density.} is a popular solution to this problem, since it allows working with interpolating operators that have suppressed overlapping with excited states. An example is provided by Gaussian smearing, which amounts to use the operator $O_g(x)$ built from the fields $\psi_g(x)$:
\begin{equation}\label{eq:pos:opsm}
\begin{split}
    & \psi_g(x)_\alpha^c =  \int dy\, \frac{e^{- (x-y)^2/2g^2}}{\sqrt{2\pi}g} \delta_{\alpha \alpha\p} \delta_{cc\p}  \psi(y)_{\alpha\p}^{c\p} \; ,
\end{split}
\end{equation}
where $c, c'$ are color indices, $\alpha, \alpha\p$ are Dirac indices and a sum is intended over $\alpha\p, c\p$. The amount of smearing, parametrised by $g$, can be tuned by looking at Fig.~\ref{fig:pos_smearing}. Without operator smearing, the smeared spectral density on the left panel grows monotonically without showing the expected peaks. Similarly, the effective mass on the right panel does not reach a plateau. When operator smearing is used, the situation improves: the effective mass does not depend on time within the statistical error after $t/a=10$, and the spectral density exhibits a peak. Due to the smearing radius of the spectral density, $\sigma/m_{\mathrm{H}}=0.8$\footnote{We use for $m_H$ the reference value provided by the effective mass analysis.}, the observed peak is the result of multiple contributions, mainly from the states $\ket{H}$ and $\ket{HH}$ which have energies $E_H = m_H$ and $E_{2H} \simeq 2m_H$. The smearing has been in fact tuned so that contributions to the spectral density are increasingly smaller after $2m_H$. For this example, we have used the interpolating operator of a pseudoscalar meson, and the configurations from the ensemble B1 of Ref.~\cite{DelDebbio:2022qgu}.

\FloatBarrier
\section{Excited states contamination}
\label{sec:pos:excited}
\begin{figure}[htb]
    \centering
    \includegraphics[width=\textwidth]{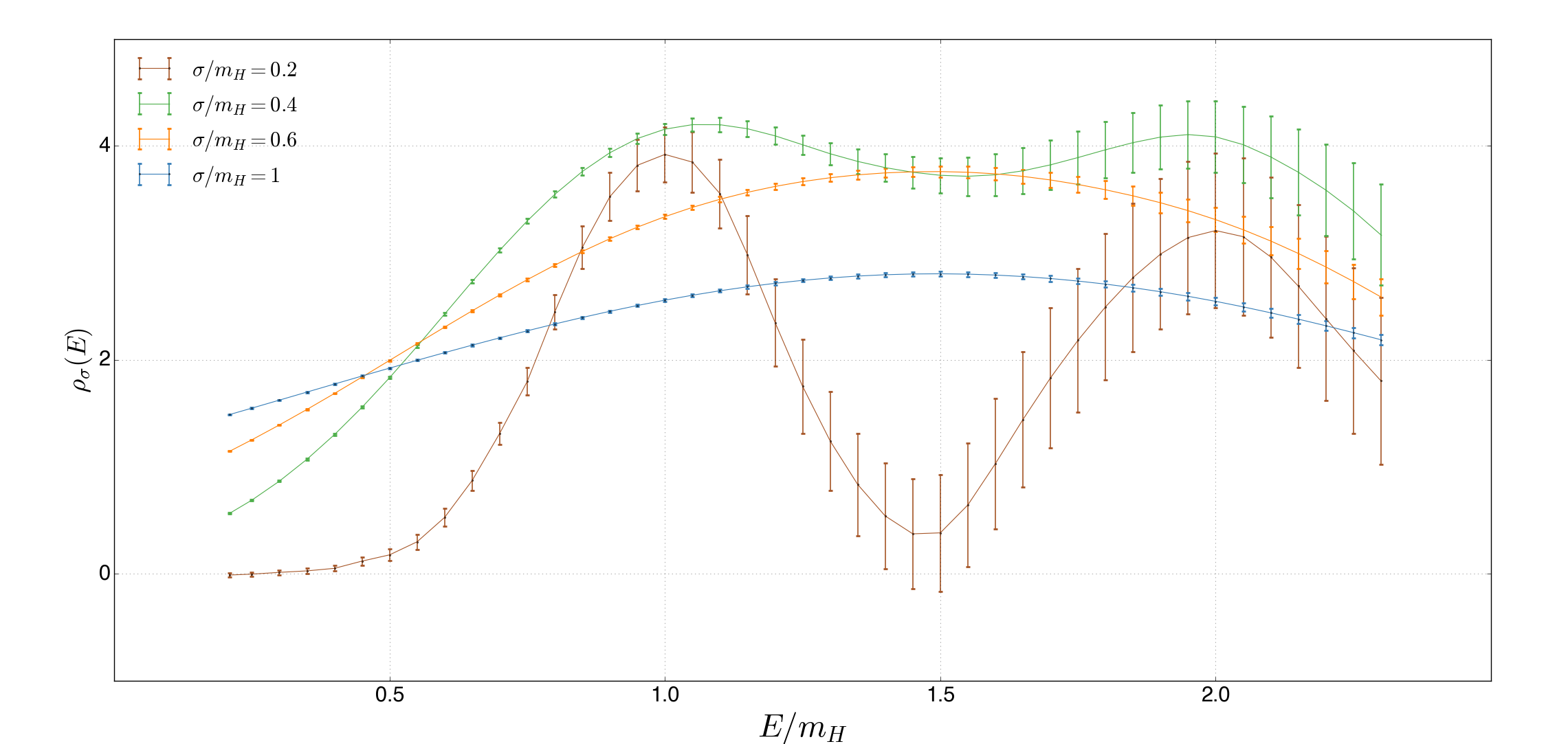}
    \caption{Smeared spectral density from synthetic data. The true spectral density has two peaks of equal height at $E=m_H$ and $E=2m_H$. If the smearing radius is too large, these cannot be distinguished. By varying the smearing radius, the peak shifts, proving that the spectral density is the result of more than one contribution. At $\sigma/m_H=0.2$ the peaks are resolved.}
    \label{fig:shift}
\end{figure}

When the smeared spectral density displays a peak, it is important to understand whether it takes contributions from a single or multiple states. This can be done by checking the location of the peak while changing the smearing radius. A signal that is not contaminated by excited states, will show a stable peak when the radius of the smearing kernel $\sigma$ is reduced. Conversely, if the peak is the result of multiple contributions, it will shift by varying $\sigma$. In this way, excited states can be detected even if the smearing radius does not allow separating them explicitly. An example is shown in Fig.~\ref{fig:shift}, in which we have used synthetic data having two states of energies $m_H$ and $2m_H$, and a relative error of $2\%$. While at $\sigma/m_H = 0.2$ the different peaks can be seen explicitly, it is also clear at larger radii ($\sigma/m_H \geq 0.6 $) that the spectral density is the result of more than one contribution, since it shifts towards $m_H$ as $\sigma$ is decreased.
\section{Fits of smeared spectral densities}
\label{sec:pos:fits}
\begin{figure}[htb]
    \centering
    \includegraphics[width=\textwidth]{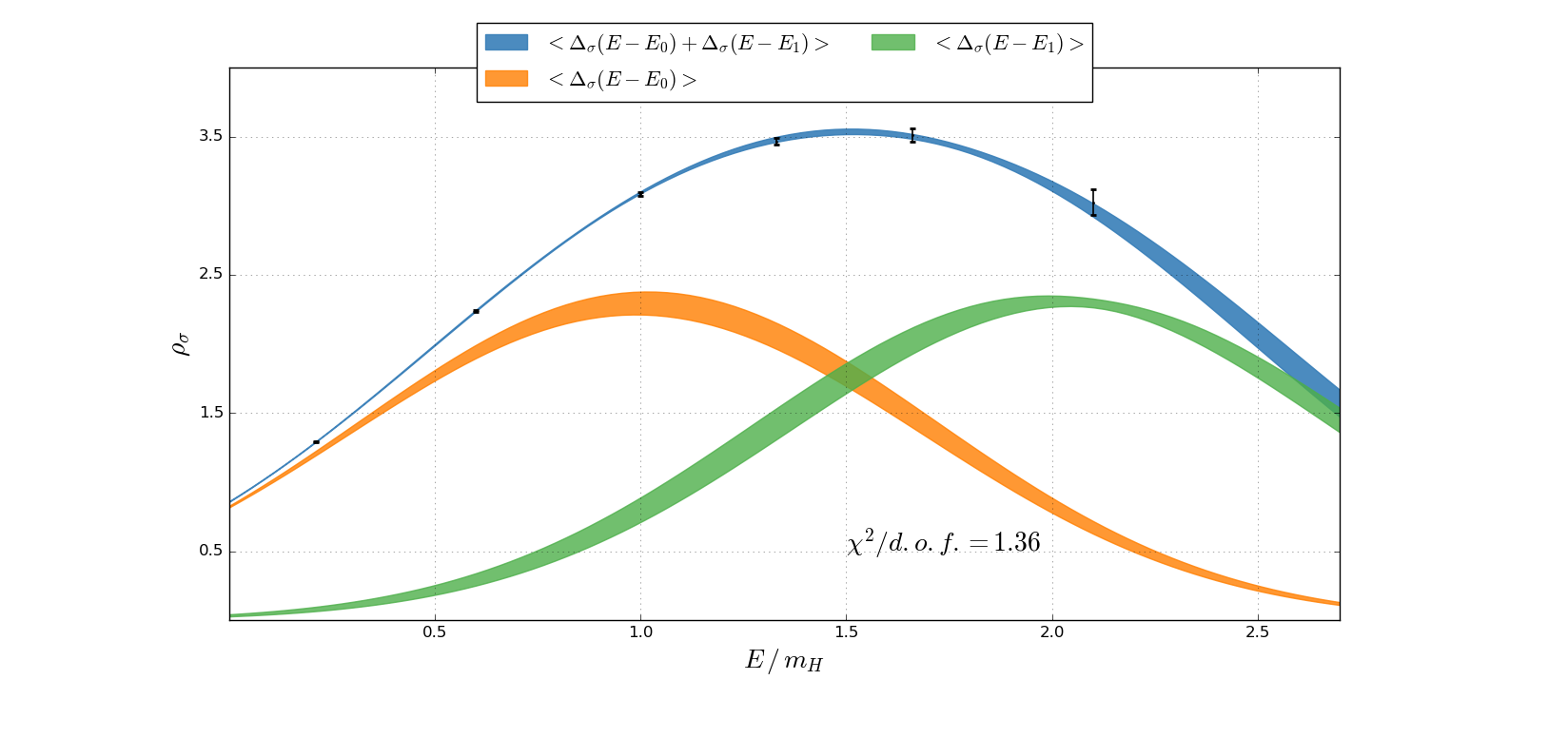}
    \caption{Example of correlated fit of a smeared spectral density (black points) from synthetic data. Despite the smearing radius being too large to separate the two contributions in the spectral reconstruction, the fit is able to identify them with great precision. The fit parameters are determined with relative uncertainty of order $1-3\%$.}
    \label{fig:fits07}
\end{figure}

A finite-volume spectral density smeared with a Gaussian kernel can be fitted to $f^{(k)}_\sigma(E)$, where the integer $k$ denotes how many states are included 
in our model function:
\begin{equation}
\label{eq:fitmodels}
    f^{(k)}_\sigma(E) = \sum_{n=1}^k w_n \Delta_\sigma(E-E_n) \; .
\end{equation}
The parameters are estimated by minimising the following $\chi^2$
\begin{equation}\label{eq:rhochisq}
    \chi_{f_\sigma^{(k)}}^2 = 
    \sum_{E,E'} \left( f_\sigma^{(k)}(E) - \rho_\sigma(E) \right)
    \text{Cov}^{-1}_{EE'}[\rho_\sigma]\left( f_\sigma^{(k)}(E') - 
    \rho_\sigma(E') \right)\; .
\end{equation}
The parameters $E_n$ and $w_n$ are related to the finite-volume energies and matrix elements according to Eq.~\eqref{eq:smeared_spectraldens_explicit}. These fits have been applied to lattice data in Ref.~\cite{DelDebbio:2022qgu}. In this section, we use synthetic data in order to show the fit results against the true values. The input data has two states with energies $m_H$ and $2m_H$, and a relative error of $2\%$. Fig.~\ref{fig:fits07} shows in the blue band an example of correlated fit to two Gaussians. The points at which the spectral densities have been evaluated are chosen in order to minimise the condition number of the covariance matrix $\text{Cov}^{-1}_{EE'}[\rho_\sigma]$ appearing in the $\chi_{f_\sigma^{(2)}}^2$ of Eq.~\eqref{eq:rhochisq}. The smearing radius of the spectral density is $\sigma/m_H = 0.7$. While this value is too large to separate the peaks, the fit is able to identify both energies. Fig.~\ref{fig:fits07} also shows the contributions coming from each Gaussian $\Delta_\sigma(E-E_n)$. The first Gaussian, and its parameters $E_0, w_0$, are mainly determined by the points at low energies, which tend to be the most precise in the reconstruction. The errors on the single Gaussians are larger than the error on their sum. For this fit, the reduced $\chi_{f_\sigma^{(2)}}^2$ is 1.36. The fit results are $E_0=1.00(1)$, $E_2=2.02(5)$, $w_0=2.30(8)$, $w_1=2.32(4)$, all in units of $m_H$.

\FloatBarrier

\section{Conclusion}
In this proceeding, we have briefly discussed how smeared spectral densities can be used at a fixed, finite volume. An example is the tuning of the overlap of a given interpolating operator with excited states, as shown in Sec.~\ref{sec:pos:smear}. Moreover, the dependence on the smearing radius $\sigma$ can be helpful in order to detect the presence of excited states in a given signal, as shown by Fig.~\ref{fig:shift}. Finally, fits of spectral densities can be used to extract finite-volume energies and matrix elements. These technologies have been applied in Ref.~\cite{DelDebbio:2022qgu} for the extraction of the ground state, in the context of composite Higgs models. Given the positive results, an interesting perspective is the application of these methods to challenging situations such as the study of resonances and baryons.
\section*{Acknowledgements}
AL and LDD received funding from the European Research Council (ERC) under the European Union’s Horizon 2020 research and innovation programme under grant agreement No 813942.  LDD is supported by the UK Science and Technology Facility Council (STFC) grant ST/P000630/1.
\FloatBarrier
\bibliographystyle{unsrt}         
\bibliography{pos22}

\begin{thebibliography}{10}

\bibitem{Maiani:1990ca}
L.~Maiani and M.~Testa.
\newblock {Final state interactions from Euclidean correlation functions}.
\newblock {\em Phys. Lett. B}, 245:585--590, 1990.

\bibitem{Luscher:1985dn}
M.~Lüscher.
\newblock {Volume Dependence of the Energy Spectrum in Massive Quantum Field
  Theories. 1. Stable Particle States}.
\newblock {\em Commun. Math. Phys.}, 104:177, 1986.

\bibitem{Luscher:1986pf}
M.~Lüscher.
\newblock {Volume Dependence of the Energy Spectrum in Massive Quantum Field
  Theories. 2. Scattering States}.
\newblock {\em Commun. Math. Phys.}, 105:153--188, 1986.

\bibitem{Kim:2005gf}
C.~h. Kim, C.~T. Sachrajda, and Stephen~R. Sharpe.
\newblock {Finite-volume effects for two-hadron states in moving frames}.
\newblock {\em Nucl. Phys. B}, 727:218--243, 2005.

\bibitem{10.1111/j.1365-246X.1968.tb00216.x}
George Backus and Freeman Gilbert.
\newblock {The Resolving Power of Gross Earth Data}.
\newblock {\em Geophysical Journal International}, 16(2):169--205, 10 1968.

\bibitem{Hansen:2019idp}
Martin Hansen, Alessandro Lupo, and Nazario Tantalo.
\newblock {Extraction of spectral densities from lattice correlators}.
\newblock {\em Phys. Rev. D}, 99(9):094508, 2019.

\bibitem{Hansen:2017mnd}
Maxwell~T. Hansen, Harvey~B. Meyer, and Daniel Robaina.
\newblock {From deep inelastic scattering to heavy-flavor semileptonic decays:
  Total rates into multihadron final states from lattice QCD}.
\newblock {\em Phys. Rev. D}, 96(9):094513, 2017.

\bibitem{Gambino:2020crt}
Paolo Gambino and Shoji Hashimoto.
\newblock {Inclusive Semileptonic Decays from Lattice QCD}.
\newblock {\em Phys. Rev. Lett.}, 125(3):032001, 2020.

\bibitem{Bulava:2021fre}
John Bulava, Maxwell~T. Hansen, Michael~W. Hansen, Agostino Patella, and
  Nazario Tantalo.
\newblock {Inclusive rates from smeared spectral densities in the
  two-dimensional O(3) non-linear \ensuremath{\sigma}-model}.
\newblock {\em JHEP}, 07:034, 2022.

\bibitem{Gambino:2022dvu}
Paolo Gambino, Shoji Hashimoto, Sandro M\"achler, Marco Panero, Francesco
  Sanfilippo, Silvano Simula, Antonio Smecca, and Nazario Tantalo.
\newblock {Lattice QCD study of inclusive semileptonic decays of heavy mesons}.
\newblock {\em JHEP}, 07:083, 2022.

\bibitem{Bulava:2019kbi}
John Bulava and Maxwell~T. Hansen.
\newblock {Scattering amplitudes from finite-volume spectral functions}.
\newblock {\em Phys. Rev. D}, 100(3):034521, 2019.

\bibitem{Bruno:2020kyl}
Mattia Bruno and Maxwell~T. Hansen.
\newblock {Variations on the Maiani-Testa approach and the inverse problem}.
\newblock {\em JHEP}, 06:043, 2021.

\bibitem{Bailas:2020qmv}
Gabriela Bailas, Shoji Hashimoto, and Tsutomu Ishikawa.
\newblock {Reconstruction of smeared spectral function from Euclidean
  correlation functions}.
\newblock {\em PTEP}, 2020(4):043B07, 2020.

\bibitem{DelDebbio:2022qgu}
Luigi Del~Debbio, Alessandro Lupo, Marco Panero, and Nazario Tantalo.
\newblock {Multi-Representation Dynamics of SU(4) Composite Higgs Models:
  Chiral Limit and Spectral Reconstructions}.
\newblock {\em arXiv:2211.09581}, 11 2022.

\end{thebibliography}
\end{document}